\documentclass[conference]{IEEEtran}
\IEEEoverridecommandlockouts

\usepackage{cite,amsmath,amssymb,graphicx,booktabs,url}
\usepackage[hidelinks]{hyperref}
\hypersetup{
  pdftitle={In RAG We Trust? Measuring Robustness of Retrieval-Augmented Generation Under Post-Retrieval Context Tampering},
  pdfauthor={Iliano Fasolino},
  pdfsubject={Retrieval-augmented generation, adversarial robustness, context tampering},
  pdfkeywords={retrieval-augmented generation, RAG, large language models, adversarial robustness, context tampering, fact verification}
}

\graphicspath{{figures/}}

\begin{document}

\title{In RAG We Trust? Measuring Robustness of Retrieval-Augmented Generation Under Post-Retrieval Context Tampering}

\author{\IEEEauthorblockN{Iliano Fasolino}
\IEEEauthorblockA{Department of Computer Science\\
University of Milan\\
Milan, Italy}}

\maketitle

\begin{abstract}
Retrieval-augmented generation grounds a language model in retrieved documents,
and that grounding is also an attack surface: an adversary who edits retrieved
text, without touching the model or the retriever, can feed the model fluent
falsehoods. We measure this on a 4-bit Llama 3.1 8B system with a MiniLM
retriever over FEVER evidence. This version replaces the measurement rather
than reinterpreting it. The earlier sweep contained an off-by-one that made a
zero-corruption cell impossible, so the row published as a clean baseline
carried one corrupted passage; a seeded re-run of 1{,}911 generations supplies a
real control, logs how many passages each corruption rule actually modified,
and adds closed-book and irrelevant-context arms. Three results follow. Scored
on corruption actually applied, the fooled rate runs 0.4\%, 35.7\%, 54.0\% and
72.0\% for zero through three corrupted passages, the zero cell being an
internal control with no adoption at all. The rules do not deliver the
corruption they request: asked for three passages, numeric substitution
achieves it in 6\% of cells, so the three rules were never comparable and we
withdraw the ranking we previously published. Controls reframe the rest, since
the generator is more accurate with no context (83.7\%) than with clean
retrieved context (77.6\%), and with irrelevant context it abstains on every
question rather than using knowledge it demonstrably has.
\end{abstract}


\section{Introduction}

Large language models answer fluently but carry two known liabilities: their
knowledge is frozen at training time, and they sometimes state false facts with
confidence. Retrieval-augmented generation \cite{lewis2020} addresses both by
fetching documents at query time and placing them in the prompt as evidence,
with the instruction to answer from that evidence rather than from memory.

The same design creates a dependency. A RAG answer is only as trustworthy as
the text that reaches the prompt. Shi et al.\ \cite{shi2023} showed that merely
irrelevant context pulls a model off track. Corrupted but on-topic context is a
stronger lever, because it wears the appearance of the evidence the model was
told to trust. Recent work on retrieval corruption shows that a small number of
crafted passages can steer a RAG system toward an attacker's answer
\cite{zou2024}, that text reaching a model through a retrieval or tool channel
is an injection surface in deployed systems \cite{greshake2023}, and that
poisoning the corpora behind such systems is practical at scale
\cite{carlini2023}.

What is still thin is a plain measurement on the kind of system a practitioner
runs on modest hardware: how far accuracy moves as corrupted context grows,
whether the type of corruption matters, and whether a failing model fabricates
or refuses. This paper reports such a measurement twice, because the first
attempt was instrumented badly enough that the second is a replacement rather
than a revision.

\textbf{Contributions.} We rebuild the measurement around a seeded harness that
expresses the corruption level as a count of passages, asserts that the two
prompts are byte-identical when that count is zero, randomises which arm is
generated first, and logs the edits actually applied to each passage. On 1{,}911
generations this yields: a dose response scored on corruption actually applied,
with a genuine zero-dose control; an edit-success rate showing that the three
corruption rules deliver very different amounts of corruption for the same
nominal level; a direct measure of poison adoption, separated from correctness;
and closed-book and irrelevant-context controls establishing what retrieval was
worth in the first place. We also state what the earlier analysis got wrong and
which of its claims we withdraw.

\section{Threat Model}
\label{sec:threat}

We fix the adversary before describing the system, because the scope of the
measurement follows from it.

The adversary can rewrite the text of passages after retrieval and before the
prompt is assembled. It cannot retrain or fine-tune the generator, cannot alter
the embedding model or the index, and cannot change which passages are
retrieved. This is the tampered-context setting: a compromised cache or serving
layer between retriever and prompt, a document store edited in place, or a
rendering step that rewrites passage text.

We do not model index poisoning, where an attacker inserts documents so that
they are retrieved at all. That threat is real \cite{zou2024,carlini2023} but it
turns on retrieval success, which this study does not manipulate. Claims here
concern what a model does with corrupted evidence once that evidence is in front
of it.

\section{Method}

\subsection{Retrieval and Generation}

Questions and passages are embedded with all-MiniLM-L6-v2 \cite{reimers2019}
into 384-dimensional vectors, normalised to unit length and stored in a FAISS
flat inner-product index \cite{johnson2019}, so the inner product equals cosine
similarity:
\begin{equation}
\mathrm{sim}(q,d)=\frac{\mathbf{e}_q\cdot\mathbf{e}_d}{\lVert\mathbf{e}_q\rVert\,\lVert\mathbf{e}_d\rVert}.
\label{eq:cos}
\end{equation}
The index is scanned exhaustively. The top $k=3$ passages go to the generator,
Llama 3.1 8B Instruct quantized to 4 bits (GGUF Q4\_K\_M) under llama.cpp on
CPU, which fits the 8\,GB machine used throughout. Decoding uses temperature
0.1 with an explicit per-generation seed. The prompt, reproduced in
Section~\ref{sec:repro}, instructs the model to answer only from the supplied
context and to reply ``Not enough information'' otherwise.

\subsection{Corruption Rules and Dose}

A corruption rewrites a retrieved passage so that it asserts something false
while staying fluent and on topic. \emph{Entity swap} replaces named entities
with same-type alternatives, turning ``the Eiffel Tower is located in Paris,
France'' into ``\ldots\ in Madrid, Spain''. \emph{Number swap} perturbs a
numeric value, turning ``over 13{,}000 miles'' into ``over 3{,}000 miles''.
\emph{Negation} inserts a negation, so ``Barack Obama was the 44th President''
becomes ``\ldots\ was never the 44th President''. These are rule-based edits
over a fixed substitution table, not model-generated adversarial text.

The corruption level is the \emph{number of passages to corrupt}, $m \in \{0,
1, 2, 3\}$, passed to the routine as a count. The earlier sweep passed a ratio
$p$ and converted it with $m=\max(1,\lfloor kp \rfloor)$, a formula that never
returns zero and maps both $p=0$ and $p=1/3$ to $m=1$;
Section~\ref{sec:artifact} reports the damage.

A requested count is not a delivered one. A rule can only corrupt a passage in
which it finds something to change, and many FEVER evidence sentences contain
no swappable numeral or no entity in the substitution table. The harness
therefore records, per passage, whether an edit was applied, and we report the
resulting \emph{applied dose} alongside the requested one.

\subsection{Outcome Labels}

Each run pairs one clean answer with one corrupted answer over the same
retrieved list, under seeds fixed by cell coordinates. Four labels are recorded.

\emph{Correctness.} The original criterion was presence of a reference keyword.
That criterion is not neutral across rules: negation leaves the keyword in
place, entity swap removes it by construction, and an answer that denies the
fact (``Madrid, not Paris'') still contains the keyword. We use a
\emph{polarity-aware} criterion that resolves abstention first, reads the yes/no
stance for boolean questions, and requires the keyword to fall outside the scope
of a negation cue.

\emph{Fooled rate.} Runs where the clean answer was correct and the corrupted
answer was not, divided by the runs that were correct before the attack. The
earlier version divided by all runs while describing it as the former.

\emph{Abstention.} Answers that decline to commit, matched against the fixed
phrase list in Section~\ref{sec:repro}. Abstention and fooled are not disjoint,
so stacked figures use a resolved three-way outcome on clean-correct runs only.

\emph{Poison adoption.} Whether the answer contains a value introduced by the
corruption, tested against the edits actually logged for that run. Adoption is a
different axis from correctness: an answer can abstain and still quote the
corrupted value. Earlier versions used a lexical-overlap proxy instead,
\begin{equation}
\mathrm{lowovl} = \mathbf{1}\!\left[\frac{\lvert W_a \cap W_d\rvert}{\lvert W_a\rvert} < 0.3\right],
\label{eq:hallu}
\end{equation}
which has two defects we quantify in Section~\ref{sec:withdrawn}: $W_d$ was the
arm's own, hence corrupted, passages, so an answer that copied the poison scored
as well supported; and abstentions were excluded before evaluation and counted
as supported, so the measure fell mechanically as abstention rose.

\section{Experimental Design}

The evaluation set holds 49 indexed query slots drawn from 37 distinct question
texts. The 12 repeats exist because the set was assembled by listing questions
per rule and then appending a mixed block that re-used earlier questions. They
are not independent observations, and every interval in this paper resamples the
37 distinct texts.

Ten information-rich passages covering the queried facts were written by the
author and added to the index, because raw FEVER evidence sentences are often
too terse to support a full answer. Retrieval diagnostics were run against that
injected map (script \texttt{retrieval\_diag.py}; reconstructed clean contexts
shipped as \texttt{clean\_contexts.json}), confirming that the top-$k$ set
usually includes a topic-relevant injected passage and that a substantial share
of retrieved text is author-written rather than raw FEVER evidence. Hard
recall@3 percentages are not restated here because the printed summary was not
reproducible from the shipped package alone. Qualitative reading is unchanged:
what follows is not mainly retrieval failure. The misses cluster on one topic,
and one query, ``How many escalators does Burj Khalifa have?'', has no
supporting passage in the index at all.

Two sweeps are referenced. The \emph{published sweep} crossed 49 slots with
three rules and four nominal levels for 588 runs; its analysis is corrected in
Sections~\ref{sec:artifact} and \ref{sec:withdrawn}. The \emph{seeded sweep}
supplies the results of this paper: three rules by three requested doses by 49
queries by three independent corruption draws, 1{,}323 grid cells, plus 147 runs
each of a clean arm, a zero-dose control, a closed-book arm and an
irrelevant-context arm, for 1{,}911 generations and about 38 hours of CPU.

\section{Results}

\subsection{The Published Sweep Had No Zero Dose}
\label{sec:artifact}

Under $m=\max(1,\lfloor kp\rfloor)$ with $k=3$, the four nominal levels realise
one, one, two and three corrupted passages. The published no-attack baseline was
an attacked cell and the first two levels were a single dose.

The data carry the signature. At nominal $0/3$ the corrupted arm quotes content
that exists only in corrupted text: ``The context states the Eiffel Tower is
located in Madrid, Germany''; ``according to statement 2, the Eiffel Tower is
located in Madrid, Spain, not France''. Correctness at that cell is 115 clean
against 102 corrupted, with discordant counts $b=14$, $c=1$ and exact McNemar
$p=9.8\times10^{-4}$. A reviewer of the previous version judged that asymmetry
too large for symmetric decoding noise and asked for an artifact search. This is
the artifact.

\subsection{What a Real Zero Dose Looks Like}

The seeded harness makes the zero cell real: it asserts the two prompts are
byte-identical before generating and randomises which arm runs first. Across 147
control runs the assertion holds in 100\% of cases and no passage is edited.

The two generations then agree verbatim in 88.4\% of runs, so decoding at
temperature 0.1 is not deterministic. What matters is that the disagreement is
symmetric: correctness splits 115 against 114, with discordant counts $b=1$,
$c=0$ and exact McNemar $p=1$. Which arm is generated first does not matter
either, at 89.7\% agreement against 87.0\%. Decoding noise of this shape cannot
manufacture the $b=14$, $c=1$ asymmetry of Section~\ref{sec:artifact}, because
it has no direction. The corruption does.

\subsection{The Rules Do Not Deliver the Dose They Request}

\begin{table}[t]
\caption{Passages actually corrupted for each requested count, seeded sweep.
A rule can only edit a passage in which it finds a target.}
\label{tab:edits}
\centering
\begin{tabular}{lrrrrrr}
\toprule
& \multicolumn{2}{c}{1 requested} & \multicolumn{2}{c}{2 requested} & \multicolumn{2}{c}{3 requested} \\
\cmidrule(lr){2-3}\cmidrule(lr){4-5}\cmidrule(lr){6-7}
Rule & mean & exact & mean & exact & mean & exact \\
\midrule
Entity swap & 0.52 & 52\% & 1.06 & 40\% & 1.59 & 29\% \\
Number swap & 0.41 & 41\% & 0.91 & 18\% & 1.31 & \textbf{6\%} \\
Negation & 0.57 & 57\% & 1.26 & 44\% & 1.92 & 37\% \\
\bottomrule
\end{tabular}
\end{table}

\begin{figure}[t]
\centering
\includegraphics[width=\columnwidth]{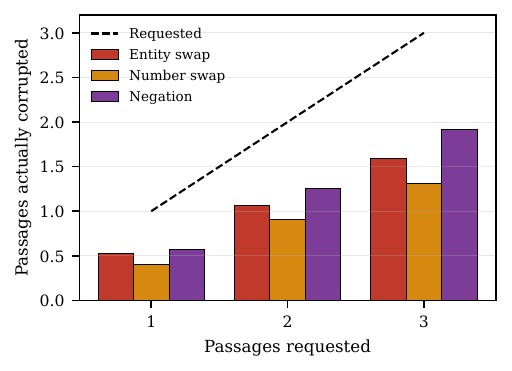}
\caption{Requested against applied corruption. The dashed line is what was
asked for. Numeric substitution falls furthest short, and the shortfall grows
with the requested dose.}
\label{fig:edits}
\end{figure}

Table~\ref{tab:edits} and Figure~\ref{fig:edits} report the gap. Asked to
corrupt all three passages, numeric substitution succeeds in 6\% of cells,
because most FEVER evidence sentences carry no swappable numeral. Negation, which
only needs a copula, manages 37\%.

Two consequences follow, and both are consequential for the earlier analysis.
Any dose response indexed on the requested level is attenuated by an unknown and
rule-dependent amount. And the three rules were never delivering comparable
corruption, so comparing their strength at equal nominal dose compares different
attacks.

\subsection{Dose Response on Corruption Actually Applied}

\begin{table}[t]
\caption{Seeded grid scored on the dose actually applied. The zero row is an
internal control: a corruption was requested but nothing could be modified, so
the model saw clean context under otherwise identical conditions.}
\label{tab:dose}
\centering
\begin{tabular}{rrrlrr}
\toprule
Applied & $n$ & Fooled & 95\% CI & Adoption & Abstain \\
\midrule
0 & 418 & 0.4 & $[0.0,\ 1.3]$ & 0.0 & 31.3 \\
1 & 511 & 35.7 & $[25.0,\ 47.6]$ & 10.8 & 34.8 \\
2 & 289 & 54.0 & $[40.1,\ 66.5]$ & 23.5 & 46.0 \\
3 & 105 & 72.0 & $[51.5,\ 92.0]$ & 15.2 & 62.9 \\
\bottomrule
\end{tabular}
\end{table}

Table~\ref{tab:dose} is the central result. On a clean-correct base the fooled
rate rises from 0.4\% with nothing modified to 72.0\% with three passages
modified, and the interval at zero excludes every other row.

The zero row deserves emphasis because the published sweep had no equivalent.
These are 418 cells in which the harness requested a corruption, the rule found
nothing to edit, and the model therefore saw clean context while every other
condition was held fixed. The fooled rate there is 0.4\% and adoption is
exactly zero. That is the behaviour a no-attack cell should show, and it is what
the published $0/3$ row failed to show.

\begin{figure*}[t]
\centering
\includegraphics[width=\textwidth]{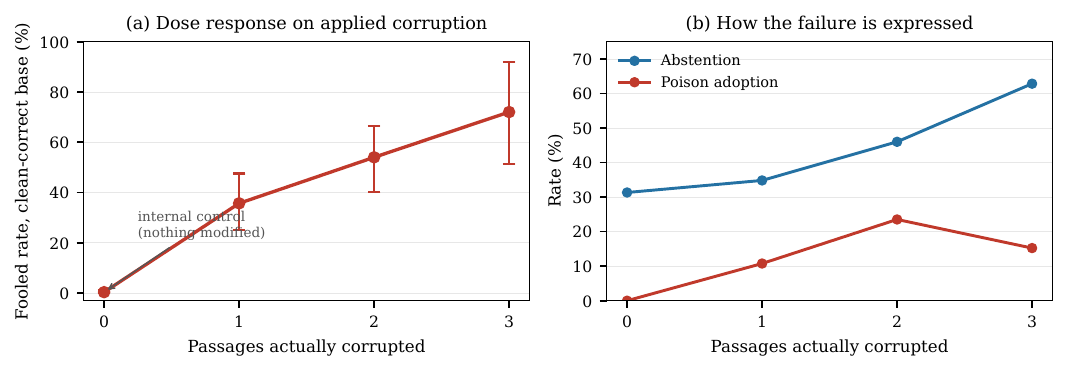}
\caption{Panel (a): fooled rate against passages actually corrupted, with
query-level bootstrap intervals. The leftmost point is the internal control.
Panel (b): abstention and poison adoption over the same axis. Abstention rises
throughout; adoption rises then falls as refusal takes over at full corruption.}
\label{fig:dose}
\end{figure*}

Figure~\ref{fig:dose} adds how the failure is expressed. Abstention climbs
monotonically, from 31.3\% to 62.9\%. Adoption rises from zero to 23.5\% at two
corrupted passages and then falls back to 15.2\% at three, which is consistent
with refusal displacing commitment once the context is wholly corrupted: the
model has less and less that it is willing to assert, including the poison.

\subsection{The Rules Are Still Not Separable}

\begin{table}[t]
\caption{Fooled rate by rule over three corruption draws, and pairwise
differences paired by query text. No interval excludes zero.}
\label{tab:rules}
\centering
\begin{tabular}{lrl}
\toprule
Rule & Fooled (\%) & 95\% CI \\
\midrule
Entity swap & 40.1 & $[28.3,\ 51.1]$ \\
Number swap & 37.4 & $[24.6,\ 50.5]$ \\
Negation & 26.6 & $[15.0,\ 39.8]$ \\
\midrule
Entity $-$ Number & $+1.1$ & $[-11.0,\ +12.5]$ \\
Entity $-$ Negation & $+7.5$ & $[-0.5,\ +16.3]$ \\
Number $-$ Negation & $+6.4$ & $[-7.1,\ +19.9]$ \\
\bottomrule
\end{tabular}
\end{table}

The previous version reported that entity swap flips the most answers and
negation is the most resilient. Table~\ref{tab:rules} does not support it, now
on three independent corruption draws rather than one. Point estimates preserve
the order, every pairwise interval contains zero, and the closest contrast,
entity against negation, has a lower bound of $-0.5$ points.

The comparison is in fact worse posed than that. Table~\ref{tab:edits} shows the
rules deliver different amounts of corruption, so a fair test would hold applied
dose equal. Restricting to cells where a rule landed all three edits leaves no
query answered that way by both entity and number substitution, and only five
shared between entity swap and negation. At this corpus the rules cannot be
compared at equal delivered corruption, because they cannot deliver equal
corruption. We withdraw the ranking and do not replace it.

\subsection{Adoption Separates the Rules Where Fooled Rate Cannot}

\begin{table}[t]
\caption{Poison adoption in the seeded sweep, tested against the edits actually
applied.}
\label{tab:adopt}
\centering
\begin{tabular}{lrrrr}
\toprule
Rule & Overall & \multicolumn{3}{c}{By applied dose} \\
\cmidrule(lr){3-5}
& & 1 & 2 & 3 \\
\midrule
Entity swap & 15.2\% & 16\% & 38\% & 14\% \\
Number swap & 15.2\% & 16\% & 32\% & 100\% \\
Negation & \textbf{1.1\%} & 1\% & 3\% & 2\% \\
\bottomrule
\end{tabular}
\end{table}

\begin{figure*}[t]
\centering
\includegraphics[width=\textwidth]{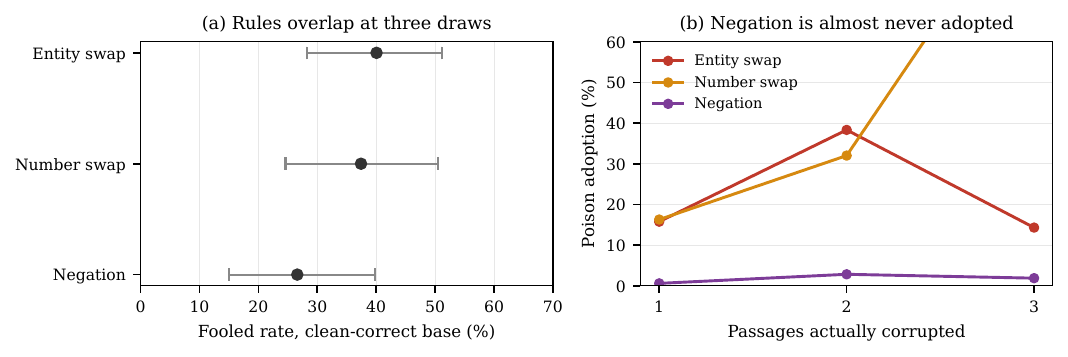}
\caption{Panel (a): fooled rate by rule with bootstrap intervals; the three
overlap. Panel (b): poison adoption over applied dose, where negation separates
sharply from the other two.}
\label{fig:rules}
\end{figure*}

Where the fooled rate cannot tell the rules apart, adoption can.
Table~\ref{tab:adopt} and Figure~\ref{fig:rules}(b) show entity and numeric
substitution adopted at 15.2\% each, against 1.1\% for negation. The model
repeats a substituted place or figure readily and almost never repeats an
inserted negation. This is the one behavioural difference between the rules that
this study establishes, and it is invisible to a correctness-based comparison.

Two further observations bear on how the earlier metrics were read. Adoption and
abstention are not exclusive: 57.6\% of adopting answers also decline to commit,
quoting the corrupted value and then reporting the context insufficient. Refusal
is therefore not evidence that corruption was rejected. And in the seeded sweep
no adopting answer was scored correct, so adoption is cleanly distinguishable
from the fooled rate rather than a relabelling of it.

\subsection{What the Context Was Worth}
\label{sec:controls}

\begin{table}[t]
\caption{Controls, 147 runs each. The closed-book arm receives no context and is
told to answer from its own knowledge; the irrelevant arm receives three true but
off-topic passages under the standard instruction.}
\label{tab:controls}
\centering
\begin{tabular}{lrr}
\toprule
Arm & Correct (\%) & Abstain (\%) \\
\midrule
Clean context & 77.6 & 12.2 \\
Closed book, no context & \textbf{83.7} & 0.0 \\
Irrelevant context & 0.0 & 100.0 \\
\bottomrule
\end{tabular}
\end{table}

\begin{figure}[t]
\centering
\includegraphics[width=\columnwidth]{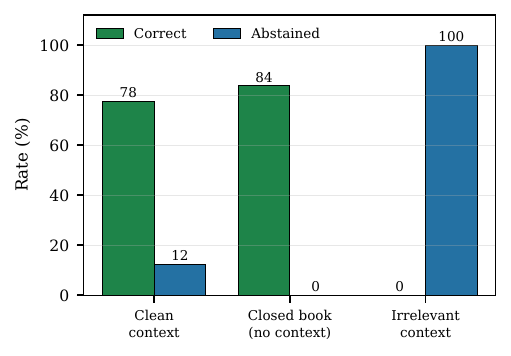}
\caption{Controls. Removing the context raises accuracy; replacing it with true
but irrelevant text collapses the model to universal abstention rather than to
its own knowledge.}
\label{fig:controls}
\end{figure}

Neither earlier version established what retrieval contributed.
Table~\ref{tab:controls} and Figure~\ref{fig:controls} settle it, and the answer
is uncomfortable: the generator is more accurate with no context at all than
with the clean retrieved context, by about six points. Eight questions are
answered correctly closed-book and lost under clean retrieval, and they are the
questions where retrieval misses. ``What team did Kaepernick play for?'' returns
``San Francisco 49ers'' without context and ``Not enough information'' with it.

The irrelevant arm gives the mechanism. Given three true but off-topic passages
the model answers correctly zero times out of 147 and abstains in every single
run. It never falls back on knowledge it demonstrably has. The instruction to
answer only from context is obeyed absolutely, so context does not compete with
parametric knowledge, it replaces it.

That reframes the dose response. Corruption is not overriding what the model
believes; it is the only thing the model is permitted to believe. The
degradation in Table~\ref{tab:dose} is the cost of binding an instruction
following model to evidence, realised when the evidence is corrupted. It also
bounds external validity: these questions concern well-known entities the
generator already knows, and a corpus of facts outside its training data would
behave differently in ways this design cannot predict.

One caveat on comparability. The closed-book arm necessarily uses a different
system instruction, since ``answer only from the context'' is undefined with no
context, so 83.7\% is not a strictly matched condition. The direction does not
depend on the margin.

\subsection{Claims Withdrawn from the Published Analysis}
\label{sec:withdrawn}

Four statements from the previous version do not survive and are withdrawn here
rather than quietly dropped.

\emph{The zero-poison baseline.} It was not one, as
Section~\ref{sec:artifact} shows, and its residual was attributed to decoding
noise and a strict matcher. Decoding noise exists but is symmetric, so it cannot
account for that cell.

\emph{The headline drop.} Reporting 77.9\% on the clean arm against 43.5\% at
full corruption mixes a between-arm gap with the attack. On the published table
the nominal within-arm attack-only drop is 69.4\% to 43.5\%, or 25.9 points; the
rebinned applied-dose drop from realised dose~1 (66.7\%) to dose~3 (43.5\%) is
23.2 points. In the seeded sweep the corresponding contrast is the
zero-to-three row of Table~\ref{tab:dose}.

\emph{The strategy ranking.} Withdrawn in Table~\ref{tab:rules}, and further
undermined by Table~\ref{tab:edits}, which shows the rules were not delivering
comparable corruption in the first place.

\emph{The fall in unsupported generation.} Eq.~\eqref{eq:hallu} scores every
abstention as supported, and abstention is what rises under attack, so the
measure declines arithmetically. On the published table the all-answers rate
fell from 28.9\% to 12.2\% while the rate over committed answers moved only from
33.1\% to 25.4\% and was not monotone in between. The proxy is also blind to
adoption by construction, since it compares an answer against the corrupted text
it received: ``located in Madrid, Spain, not France'' overlaps the true context
at 0.67, far above the threshold. Direct adoption measurement replaces it.

A fifth claim is weakened rather than withdrawn. Abstention was offered as a
cheap warning signal. It still rises with dose, but 57.6\% of adopting answers
abstain, and a merely irrelevant context produces 100\% abstention, so the signal
does not distinguish corruption from useless retrieval and does not imply the
corruption was rejected.

\section{Related Work}

Retrieval-augmented generation was introduced to ground generation in
retrievable evidence \cite{lewis2020}, extending retrieval-integrated
pre-training \cite{guu2020}. The failure mode here is the inverse of the
intended one: grounding helps on a clean corpus and hurts on a corrupted one.

On the adversarial side, Zou et al.\ \cite{zou2024} show that injecting a small
number of crafted passages into a knowledge base steers a RAG system to
attacker-chosen answers, and frame retrieval corruption as an attack surface
distinct from prompt manipulation. Greshake et al.\ \cite{greshake2023} show
that any channel delivering text into a model's context functions as an indirect
injection vector in deployed applications. Carlini et al.\ \cite{carlini2023}
show that poisoning the web-scale corpora behind such systems is practical. Our
setting is narrower than all three: the adversary edits already-retrieved text,
which isolates the generation stage from retrieval success.

Shi et al.\ \cite{shi2023} establish that irrelevant context degrades reasoning.
Our irrelevant-context arm sharpens that for this setting: under an
answer-only-from-context instruction, irrelevant context does not degrade the
answer, it suppresses it entirely. Zhou et al.\ \cite{zhou2023} caution that
evaluation data can leak into training; we keep the query set fixed across all
cells, though we cannot rule out that FEVER-derived facts about well-known
entities are present in the generator's training data, and
Section~\ref{sec:controls} suggests they are.

\section{Limitations and Reproducibility}
\label{sec:repro}

\subsection{What This Study Does Not Establish}

The measurement is one quantized 8B model, one retriever, $k=3$, 37 distinct
questions and rule-based corruption. Absolute rates should not be read as
general. No pairwise contrast between corruption rules is separable by fooled
rate, and at this corpus the rules cannot be equalised on delivered corruption,
so we make no ranking claim.

Three gaps remain. Correctness and adoption are both automated and neither was
validated against human judgement or an entailment model on a stratified sample;
the adoption rule is sharp for substituted entities and numbers and weak for
negation, where it looks for an explicit cue rather than a flipped truth value,
so 1.1\% is a lower bound for that rule. The corruption is rule-based, and an
attacker using a model to write internally consistent passages would remove the
contradiction cues our entity swaps sometimes leave. And $k$ is fixed at 3
throughout, so this study provides no evidence about whether retrieving more
passages helps or hurts; the previous version's advice to raise $k$ alongside
source diversity was unsupported and is withdrawn.

\subsection{Reproducibility Details}

The system prompt is: \emph{``You are a fact-checking assistant. Answer the
question based ONLY on the provided context. If the context doesn't contain
enough information, say `Not enough information'.''} The user turn is the
numbered context, then \texttt{Question:}, then \texttt{Answer:}, wrapped in the
Llama 3.1 chat template. Abstention is matched, case-insensitively, against:
\emph{not enough information}, \emph{cannot determine}, \emph{not provided},
\emph{unclear}, \emph{no information}, \emph{don't know}, \emph{cannot answer}.

Versions are pinned as follows: llama-cpp-python 0.3.16, sentence-transformers
5.2.2, faiss-cpu 1.13.2, numpy 1.26.2, pandas 2.1.4; generator Llama 3.1 8B
Instruct GGUF Q4\_K\_M with SHA-256
\texttt{7b064f5842bf9532\allowbreak c91456deda288a1b\allowbreak
672397a54fa729aa\allowbreak 665952863033557c};
embedder all-MiniLM-L6-v2; corpus the FEVER dev split \cite{thorne2018},
19{,}597 deduplicated evidence sentences, plus ten author-written passages;
hardware an Apple M1 with 8\,GB, CPU inference only (\texttt{n\_gpu\_layers=0}).
Seeds are derived per generation from a checksum of the cell coordinates.
Bootstrap intervals use 10{,}000 resamples of the 37 distinct query texts with
seed 42.

One quirk applies to every run reported here. The prompt is built with an
explicit \texttt{<|begin\_of\_text|>} token and the runtime prepends its own, so
each prompt carries a duplicated leading token. It is a deviation from the
reference chat template, kept in the seeded harness so old and new runs stay
comparable, and flagged rather than silently changed.

The artifact contains the corruption code, the seeded harness, the 49 queries
with reference keywords, the ten injected passages, the index build script, both
run tables, the scoring and analysis scripts and the figure code. Release is
pending and will be linked from the version of record.

\section{Conclusion}

We measured a small quantized RAG system under post-retrieval context tampering,
replacing an earlier measurement whose instrumentation was faulty. Scored on
corruption actually applied, the fooled rate runs 0.4\%, 35.7\%, 54.0\% and
72.0\% across zero to three corrupted passages, with the zero cell an internal
control that shows no adoption at all. Abstention rises throughout. Adoption
peaks in the middle of the range and falls at full corruption as refusal
displaces commitment.

The comparative claim we previously published does not survive. The three
corruption rules cannot be separated by fooled rate on three independent draws,
and they cannot be equalised on delivered corruption either, since numeric
substitution lands its full requested dose in 6\% of cells. Where they do differ
is adoption: substituted entities and figures are repeated at 15.2\%, inserted
negations at 1.1\%.

The practical reading is narrower than before and, we think, more useful. Under
an instruction to answer only from retrieved text, that instruction is obeyed
absolutely: with off-topic context the model abstained on all 147 runs rather
than using knowledge it plainly had, and with no context at all it scored higher
than with clean retrieval. Evidence quality therefore sets a hard ceiling,
corrupted evidence has no internal competitor, and retrieval should be justified
per query set rather than assumed, because on facts a model already knows it can
subtract accuracy instead of adding it. Abstention is a weaker signal than we
previously suggested, since it accompanies adoption more often than not and is
also what useless retrieval produces.

What this design still cannot answer is how any of it behaves on facts outside
the generator's training data, which is precisely where retrieval should earn
its place, and whether model-written corruption that removes the internal
inconsistencies of rule-based edits would be adopted more readily. Both are the
next measurements rather than the next argument.


\begin{thebibliography}{00}
\bibitem{lewis2020} P. Lewis \emph{et al.}, ``Retrieval-augmented generation for
knowledge-intensive NLP tasks,'' in \emph{Advances in Neural Information Processing
Systems}, vol.\ 33, 2020, pp.\ 9459--9474.

\bibitem{thorne2018} J. Thorne, A. Vlachos, C. Christodoulopoulos, and A. Mittal,
``FEVER: a large-scale dataset for fact extraction and verification,'' in
\emph{Proc.\ Conf.\ North American Chapter of the Association for Computational
Linguistics (NAACL-HLT)}, 2018, pp.\ 809--819.

\bibitem{shi2023} F. Shi \emph{et al.}, ``Large language models can be easily
distracted by irrelevant context,'' in \emph{Proc.\ Int.\ Conf.\ Machine Learning
(ICML)}, 2023.

\bibitem{zou2024} W. Zou, R. Geng, B. Wang, and J. Jia,
``PoisonedRAG: knowledge corruption attacks to retrieval-augmented generation of
large language models,'' in \emph{34th USENIX Security Symposium (USENIX Security
25)}, 2025, pp.\ 3827--3844. Also \emph{arXiv preprint} arXiv:2402.07867, 2024.

\bibitem{greshake2023} K. Greshake, S. Abdelnabi, S. Mishra, C. Endres, T. Holz,
and M. Fritz, ``Not what you've signed up for: compromising real-world
LLM-integrated applications with indirect prompt injection,'' in \emph{Proc.\ ACM
Workshop on Artificial Intelligence and Security (AISec)}, 2023.

\bibitem{carlini2023} N. Carlini \emph{et al.}, ``Poisoning web-scale training
datasets is practical,'' \emph{arXiv preprint} arXiv:2302.10149, 2023.

\bibitem{zhou2023} K. Zhou \emph{et al.}, ``Don't make your LLM an evaluation
benchmark cheater,'' \emph{arXiv preprint} arXiv:2311.01964, 2023.

\bibitem{guu2020} K. Guu, K. Lee, Z. Tung, P. Pasupat, and M. Chang, ``REALM:
retrieval-augmented language model pre-training,'' in \emph{Proc.\ Int.\ Conf.\
Machine Learning (ICML)}, 2020, pp.\ 3929--3938.

\bibitem{reimers2019} N. Reimers and I. Gurevych, ``Sentence-BERT: sentence
embeddings using Siamese BERT-networks,'' in \emph{Proc.\ Conf.\ Empirical Methods
in Natural Language Processing (EMNLP)}, 2019, pp.\ 3982--3992.

\bibitem{johnson2019} J. Johnson, M. Douze, and H. J\'egou, ``Billion-scale
similarity search with GPUs,'' \emph{IEEE Trans.\ Big Data}, vol.\ 7, no.\ 3,
pp.\ 535--547, 2019.
\end{thebibliography}
\end{document}